\documentclass[fleqn,12pt,twoside]{article}
\usepackage[headings]{espcrc1}

\readRCS
$Id: espcrc1.tex,v 1.2 2004/02/24 11:22:11 spepping Exp $
\usepackage{graphicx}
\usepackage[figuresright]{rotating}

\usepackage{epsfig}

\newcommand{\AmS}{{\protect\the\textfont2
  A\kern-.1667em\lower.5ex\hbox{M}\kern-.125emS}}

\def\sl#1{\slash{\hspace{-0.2 truecm}#1}}

\def\be{\begin{equation}}
\def\ee{\end{equation}}
\def\bea{\begin{eqnarray}}
\def\eea{\end{eqnarray}}

\def\ket#1{\hbox{$\vert #1\rangle$}}   
\def\bra#1{\hbox{$\langle #1\vert$}}   
\def\oneh{{\textstyle {1\over 2}}}

\def          
\simeq{{\ \lower2pt\hbox{$-$}\mkern-13mu \raise2pt \hbox{$\sim$}\ }} 

\def\bold#1{\setbox0=\hbox{$#1$}%
      \kern-.02em\copy0\kern-\wd0
      \kern.04em\copy0\kern-\wd0
      \kern-.02em\raise.0433em\box0 }
\def\smbold#1{\setbox0=\hbox{$\scriptstyle#1$}%
      \kern-.02em\copy0\kern-\wd0
      \kern.04em\copy0\kern-\wd0
      \kern-.02em\raise.0433em\box0 }

\def\oggi{\number\day.\space 
\ifcase\month\or
 1.\or 2.\or 3.\or 4.\or 5.\or 6.\or
 7.\or 8.\or 9.\or 10.\or 11.\or 12.\fi
 \space\number\year}

\title{Generalized parton distributions in a meson cloud model}

\author{B. Pasquini
\address[UP]{Dipartimento di Fisica Nucleare e Teorica, Universit\`a degli
Studi di Pavia and INFN, Sezione di Pavia, Pavia, Italy}, and 
S. Boffi\addressmark[UP]}
       
\runtitle{Generalized parton distributions in a meson cloud model}
\runauthor{B. Pasquini, and S. Boffi}

\begin{document}

\maketitle

\begin{abstract}
We present a model calculation 
of the generalized parton distributions where the nucleon is described by 
a quark core surrounded by a mesonic cloud. 
In the one-meson approximation, we expand the Fock state of the 
physical nucleon in a series  involving  a bare nucleon and 
two-particle, meson-baryon, states. We discuss the role of
the different Fock-state components of the nucleon by deriving 
a convolution formalism for  the unpolarized generalized
parton distributions, and showing predictions at different kinematics.
\end{abstract}

\section{The meson-cloud model for the nucleon}
\label{wf_mcm}

%

The convolution model for the physical nucleon, where the bare nucleon is 
dressed by its virtual meson cloud, has a long and successful history in 
explaining properties such as form factors~\cite{DHSS97} and 
parton distributions~\cite{Speth98}. In this
 paper it has been revisited and applied for the first time to study 
 Generalized Parton Distributions (GPDs) that have recently been 
introduced and discussed in connection with Deeply Virtual Compton Scattering
(DVCS) 
and hard exclusive meson production (for reviews, see 
Refs.~\cite{jig,radyushkin01,goeke,markusthesis}).
%


The basic assumption of the meson-cloud model is that the state of the 
physical nucleon 
$\tilde N$ can be decomposed according to the 
meson-baryon Fock-state expansion as a superposition of a bare nucleon state 
and states containing virtual mesons associated with recoiling baryons. 
This state, with four-momentum 
$p_N^\mu=(p^-_N,p^+_N,{\bf p}_{N\perp})\equiv(p^-_N,\tilde p_N)$ and helicity $\lambda$, is an eigenstate of the light-cone Hamiltonian
\begin{equation}
H_{LC}= \sum_{B,M} \left[H_0^B(q) + H_0^M(q)+H_I(N,BM)\right].
\label{eq:1}
\end{equation}
In Eq.~(\ref{eq:1}),
$H_0^B(q)$ stands for the effective-QCD Hamiltonian which 
governs the constituent-quark dynamics, and leads to the confinement of 
three quarks in a baryon state;
analogously, $H_0^M(q)$ describes the quark interaction in a meson state,
and $H_I(N,BM)$ is the
nucleon-baryon-meson interaction,
and the sum is over all the possible baryon and meson configurations in which
the nucleon can virtually fluctuate.
In the calculation presented here we use perturbative treatment of the meson
effects. So we truncate the Fock space expansion of the nucleon state  
to the one-meson components, and  expand the nucleon wavefunction 
in terms of the eigenstates of the bare Hamiltonian 
$H_0\equiv H_0^B(q) + H_0^M(q)$.
The corresponding state of the physical nucleon
$|\tilde N\rangle$ can be written as 
\begin{eqnarray}
|\tilde p_N,\lambda;\tilde N\rangle
& &
=\sqrt{Z}|\tilde p_N,\lambda; N\rangle
+
\sum_{B,M}
\int \frac{{\rm d}y{\rm d}^2{\bf k}_{\perp}}{2(2\pi)^3}\,
\frac{1}{\sqrt{y(1-y)}}
\sum_{\lambda',\lambda''}
\phi_{\lambda'\lambda''}^{\lambda \,(N,BM)}(y,{\bf k}_\perp)\nonumber\\
& &\quad{}\times
|yp^+_N,{\bf k}_{\perp}+y{\bf p}_{N\perp},\lambda';B\rangle\,
|(1-y)p^+_N,-{\bf k}_{\perp}+(1-y){\bf p}_{N\perp},\lambda'';M\rangle,
\label{eq:14}
\end{eqnarray}
where we introduced the function 
$\phi_{\lambda'\lambda''}^{\lambda\,(N,BM)}(y,{\bf k}_\perp)$ to define the 
probability amplitude for a nucleon with helicity $\lambda$ to fluctuate into 
a virtual $BM$ system with the baryon having helicity $\lambda'$, longitudinal 
momentum fraction $y$ and transverse momentum ${\bf k}_\perp$, and the meson 
having helicity $\lambda''$, longitudinal momentum fraction $1-y$ and 
transverse momentum $
-{\bf k}_\perp$.
Furthermore, in Eq.~(\ref{eq:14}), the constant $Z,$ giving the probability
that the physical nucleon is a bare core state,
ensures the correct normalization of the nucleon wave function:
\begin{equation}
\langle p'^+,{\bf p}'_\perp, \lambda ';H\vert p^+,{\bf p}_\perp\lambda;H\rangle=
2(2\pi)^3p^+\delta(p'^+-p^+)\delta^{(2)}({\bf p}'_\perp -{\bf p}_\perp)
\delta_{\lambda\lambda'}.
\label{eq:7}
\end{equation}


\section{The unpolarized generalized parton distributions}
\label{gpds}

In the definition of GPDs we choose a symmetric frame of 
reference where the virtual photon momentum $q^\mu$ and the average nucleon 
momentum $\bar p_N^\mu=\oneh(p_N^\mu+{p'}_N^\mu)$ are collinear along 
the $z$ axis and in opposite  directions.
Furthermore, $Q^2 = -q^\mu q_\mu$ is the space-like virtuality that defines the scale of the process,  $t=\Delta^2=({p'}_N^\mu-p_N^\mu)^2$ is the invariant transferred momentum square, and the skewness $\xi$ describes the longitudinal change of the nucleon momentum,  $2\xi=-\Delta^+/\bar p^+_N$.

For each flavor $q$ the soft amplitude corresponding to unpolarized GPDs reads 
\be
\label{eq:29}
F^q_{\lambda'_N\lambda_N}(\bar x,\xi,{\bf \Delta}_\perp) =
\left. \frac{1}{2\sqrt{1-\xi^2}}
\int \frac{{\rm d}z^-}{2\pi}\, e^{i\bar x\,\bar p_N^+y^-}
\bra{p'_N,\lambda'_N}\overline\psi(-\oneh z)\,\gamma^+\, 
\psi(\oneh z)\ket{p_N,\lambda_N}
\right\vert_{z^+={\bf z}_\perp=0},
\ee
where $\bar x$ defines the fraction of the quark light-cone momentum 
(${\overline k}^+={\bar x}\,{\bar p}_N^+$), $\lambda_N$
 ($\lambda'_N$) is the helicity of the initial (final) nucleon, and 
the quark-quark correlation function is integrated along the light-cone 
distance $z^-$ at equal light-cone time ($y^+=0$) and zero transverse 
separation (${\bf z}_\perp=0$) between the quarks. The leading twist 
(twist-two) part of this amplitude can be parametrized
in terms of 
 the chiral-even helicity conserving GPD, $H^q(\bar x,\xi,{\bf \Delta}_\perp)$, 
and  the helicity flipping GPD, $E^q(\bar x,\xi,{\bf \Delta}_\perp) $, 
for partons 
of flavor $q$.

In the following we will derive the convolution formulas
for the GPDs in the three different regions
 corresponding to $\xi\leq \bar x \leq 1,$ $-\xi\leq \bar x \leq \xi,$ 
and $-1\leq \bar x \leq -\xi.$


\subsection{The region $\xi \leq \bar x\leq 1$}
\label{sub1}

In this region the GPDs describe the emission of a quark from the nucleon with
 momentum fraction $\bar x +\xi$ and its reabsorption with momentum 
fraction $\bar x-\xi$. In the meson-cloud model, the virtual photon can 
hit 
either the bare nucleon $N$ or one of the higher Fock states. As a consequence,
 the DVCS amplitude can be written as the sum of two contributions
\begin{eqnarray}
F^q_{\lambda'_N\lambda_N}(\bar x,\xi,{\bf \Delta}_\perp) =
Z\,
F^{q, bare}_{\lambda'_N\lambda_N}(\bar x,\xi,{\bf \Delta}_\perp) 
+\delta F^q_{\lambda'_N\lambda_N}(\bar x,\xi,{\bf \Delta}_\perp),
\label{eq:bmcomp}
\end{eqnarray}
where $F^{q, bare}$ is the contribution from the bare proton, described in 
terms of Fock states with three valence quarks, and $\delta F^q$ is the 
contribution from the $BM$ Fock components of the nucleon state, corresponding 
to five-parton configurations. 
The valence-quark contribution $
F^{q, bare}_{\lambda'_N\lambda_N}$
 can be calculated in  the light-front 
overlap representation derived in Ref.~\cite{DFJK}, and applied  
to the case of $N=3$ valence quarks in in Ref~\cite{BPT,BPT2}, 
where one can also find 
the explicit expression 
in terms of bare-nucleon light-cone wave functions (LCWFs)
 derived in a constituent quark model.
The $\delta F^q_{\lambda'_N\lambda_N}$ term 
in Eq.~(\ref{eq:bmcomp})
can further be split into 
two contributions, with the active quark belonging either to the 
baryon ($\delta F^{q/BM}$) or to the meson ($\delta F^{q/MB}$), 
i.e.
\begin{eqnarray}
\delta F^q_{\lambda'_N\lambda_N}(\bar x,\xi,{\bf \Delta}_\perp)=
\sum_{B,M}
\left[
\delta F^{q/BM}_{\lambda'_N\lambda_N}(\bar x,\xi,{\bf \Delta}_\perp)
+\delta F^{q/MB}_{\lambda'_N\lambda_N}(\bar x,\xi,{\bf \Delta_\perp})
\right].
\label{eq:fqb}
\end{eqnarray}
The first term 
in Eq.~(\ref{eq:fqb}) corresponds to the case when
the baryon is taken out from the initial proton with a 
fraction $\bar y_B +\xi$ of the average plus-momentum $\bar p_N^+$,
 and after the interaction with the initial and final photons
is reinserted back into the final proton with  a fraction $\bar y_B-\xi$
 of the average plus-momentum $\bar{p}_N^{\,+}$. The transverse momentum of the
 baryon is  $\bar{{\bf p}}_{B\perp}-{\bf \Delta}_\perp/2$ before, and 
$\overline{{\bf p}}_{B\perp}+{\bf \Delta}_\perp/2$ after the scattering
 process.
The meson substate is a spectator during the whole scattering process.
As final result, the convolution formula for 
$\delta F^{q/BM}_{\lambda'_N\lambda_N}$
reads
\begin{eqnarray}
\delta F^{q/BM}_{\lambda'_N\lambda_N}(\overline x,\xi,{\bf \Delta}_\perp)
&=&
\frac{1}{\sqrt{1-\xi^2}}
\sum_{M}\sum_{\lambda,\lambda',\lambda''}
\int_{\overline x}^1
\frac{{\rm d}\overline y_B}{\overline y_B}
\int
\frac{{\rm d}^2\overline{\bf p}_{B\perp}}{2(2\pi)^3}
F^{q/B}_{\lambda'\lambda}
\left(\frac{\overline x}{\overline y_B},\frac{\xi}{\overline y_B}, 
{\bf \Delta}_\perp\right)
\nonumber\\
& &{}\times
\phi^{\lambda_N\,(N,BM)}_{\lambda\lambda''}(\tilde y_B,\tilde{\bf k}_{B\perp})\,
[\phi^{\lambda'_N\,(N,BM)}_{\lambda'\lambda''}(\hat y'_B,\hat{\bf k}_{B\perp})]^*,
\label{eq:baryondglap}
\end{eqnarray}
where
$
F^{q/B}_{\lambda'\lambda}$
is the scattering amplitude from the active baryon in the $BM$ component of 
the nucleon, and can be obtained in terms of the baryon LCWFs
as explained in Ref.~\cite{pb}. 

Analogously, we can derive the meson contribution to the scattering amplitude,
 corresponding to the case when the pion takes part to the interaction 
process while the baryon remains as a spectator. 
In such a case the role of the meson and baryon substates is interchanged with 
respect to the situation described before.
The meson is taken out from the initial proton with a 
fraction $\overline y_M +\xi$ of the average plus-momentum $\overline p_N^+$,
 and after the interaction with the initial and final photons
is reinserted back into the final proton with  a fraction $\overline y_M-\xi$
 of the average plus-momentum $\overline{p}_N^{\,+}$. The transverse momentum of the
 meson is  $\overline{{\bf p}}_{M\perp}-{\bf \Delta}_\perp/2$ before, and 
$\overline{{\bf p}}_{M\perp}+{\bf \Delta}_\perp/2$ after the scattering
 process.
Viceversa, the baryon substate is inert during the whole scattering process.
Therefore the meson contribution to the 
$\delta F^q_{\lambda'_N\lambda_N}$ scattering amplitude is
 given by
\begin{eqnarray}
\delta F^{q/MB}_{\lambda'_N\lambda_N}(\overline x,\xi,{\bf \Delta}_\perp)
&=&
\frac{1}{\sqrt{1-\xi^2}}
\sum_{B}\,
\sum_{\lambda,\lambda',\lambda''}
\int_{\overline x}^1\frac{{\rm d}\overline y_M}{\overline y_M}
\int\frac{{\rm d}^2\overline{\bf p}_{M\perp}}{2(2\pi)^3}
F^{q/M}_{\lambda'\lambda}
\left(\frac{\overline x}{\overline y_M},\frac{\xi}{\overline y_M}, {\bf \Delta}_\perp\right)
\nonumber\\
& &{}\times
\phi^{\lambda_N\,(N,BM)}_{\lambda''\lambda}(1-\tilde y_M,-\tilde{\bf k}_{M\perp})\,
[\phi^{\lambda'_N\,(N,BM)}_{\lambda''\lambda'}(1-\hat y'_M,-\hat{\bf k}_{M\perp})]^*,
\label{eq:mesondglap}
\end{eqnarray}
where  
$
F^{q/M}_{\lambda'\lambda}$
is the scattering amplitude from the active meson in the $BM$ component of the 
nucleon, given explicitly in Ref.~\cite{pb} in terms of meson LCWFs. 


\subsection{The region $-1 \leq \overline x\leq -\xi$}
\label{sub2}

In this region, the scattering amplitude describes the emission of an 
antiquark from the nucleon with momentum fraction $-(\overline x+\xi)$ and 
its reabsorption with momentum fraction $-(\overline x-\xi)$. Therefore, 
the only nonvanishing contribution comes from the active antiquark in the
meson substate of the $BM$ Fock component of the nucleon wavefunction, i.e.
\begin{eqnarray}
F^q_{\lambda'_N\lambda_N}(\overline x,\xi,{\bf \Delta}_\perp)
=\delta F^{q/MB}_{\lambda'_N\lambda_N}(\overline x,\xi,{\bf \Delta}_\perp),
\end{eqnarray}
where $\delta F^{q/MB}_{\lambda'_N\lambda_N}(\overline x,\xi,{\bf \Delta}_\perp)$
is given by
 the same convolution 
formula~(\ref{eq:mesondglap}), with the integration range 
over $\overline y_M$ 
between $-\overline x$ and $1$, and with the LCWF overlap 
representation of $F^{q/M}_{\lambda'\lambda}$ in the range 
$-1 \leq \overline x\leq -\xi$.
 

\subsection{The region $-\xi \leq \overline x\leq \xi$}
\label{sub3}

In this region, the scattering amplitude describes the emission of a 
quark-antiquark pair from the initial proton. 
In the Fock-state decomposition of the initial and final nucleons we have to
 consider only terms where the initial state has the same parton content as the
 final state plus an additional quark-antiquark pair. In the present 
meson-cloud model, the initial state is given by the five-parton component of the $|BM\rangle $ Fock state, 
while the final state is described by the three-valence
 quark configuration, multiplied by the 
normalization factor $\sqrt{Z}$.
In principle, we can distinguish between two cases: $i$) 
the active quark belongs to the baryon substate, and the active 
antiquark is in the parton configuration of the meson substate of the initial
 nucleon; $ii$) both the active quark and  antiquark belong to the meson 
substate
 of the initial nucleon and the baryon is a spectator
during the scattering process.
However, this last contribution is vanishing because it involves 
the overlap of two orthogonal states, i.e. the wave 
functions of the baryon in the initial state and of the bare 
nucleon in the final state.
As a consequence, 
the only non vanishing contribution corresponds to the case $i$), and
it can be explicitly derived in terms of the overlap
between the five and three parton components of the LCWFs of the physical
 nucleon state as explained in details in Ref.~\cite{pb}.

\section{Results}
\label{results}

In this section  we present results
for the GPDs in the meson cloud model
by restricting ourselves to consider only 
the pion-cloud contribution and disregarding the contributions from mesons 
of higher masses which are suppressed.
As a consequence, the accompanying baryon in the $\ket{B\pi}$ component of
the dressed proton is a nucleon or a $\Delta.$
For the bare-hadron constituents of the nucleon state
 we use the LCWFs in the relativistic constituent model adopted in previous works to describe the valence contribution to GPDs both in the chiral-even and chiral-odd sector~\cite{BPT,BPT2,PBT,PPB,ptb,pb}.
The model assumption to describe the vertex functions 
$\phi^{\lambda(N,BM)}_{\lambda',\lambda''}$
are given in details in Ref.~\cite{pb}.

\begin{figure}[t]
\begin{center}
\epsfig{file=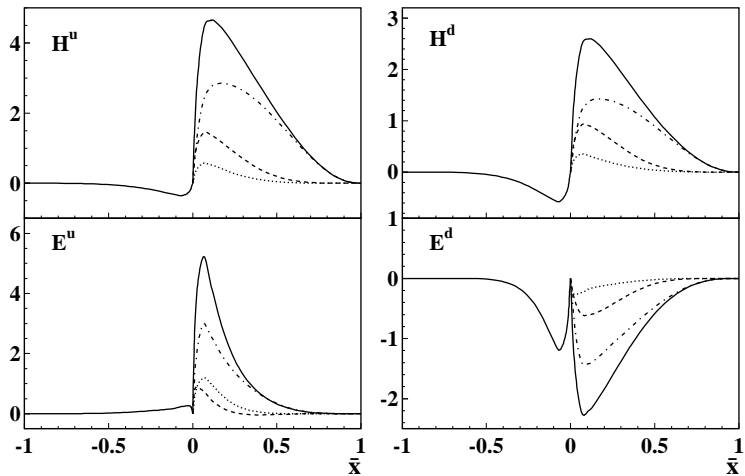,  width=10cm}
\end{center}
\caption{\small The different contributions to the spin-averaged ($H^q$, upper panels) and helicity-flip ($E^q$, lower panels) generalized parton distributions calculated in the meson-cloud model for flavours $u$ (left panels) and $d$ (right panels), at $\xi=0$ and $t=0$. Dashed lines: baryon contribution from the $\vert BM\rangle$ component. Dotted lines: meson contribution from the $\vert BM\rangle$ component.  Dashed-dotted lines: contribution from the bare nucleon. Full curves: total result as a sum of the different contributions.
}
\label{fig:fig1}
\end{figure}

First let us study the forward limit, $\xi=0$, $t=0$. 
In this limit the scattering amplitude without nucleon helicity flip reduces 
to the ordinary parton distribution, and the convolution formulas for
$H^q(\bar x,0,0)$
coincides with the formulation of the meson cloud model for the ordinary parton distributions in deep 
inelastic processes~\cite{Speth98}.
In Fig.~\ref{fig:fig1} the spin-averaged $H^q$ and the helicity-flip 
$E^q$ GPDs are plotted together with the separated contributions from 
the bare proton (dashed-dotted line), the baryon (dashed lines) and 
the meson (dotted lines) in the baryon-pion fluctuation. All these contributions add up incoherently to give the total result (full curves). The bare proton 
contribution is
always 
positive within its support ($0\le\overline x\le 1$) with the exception of 
$E^d$ for which it is negative. The same behaviour characterizes the baryon 
contribution from the baryon-pion fluctuation that is also limited to the 
range $0\le\overline x\le 1$, consistently with the assumption that the 
only active degrees of freedom for such a baryon are the valence quarks. 
The sea-quark contribution, extending all over the full range 
$-1\le\overline x\le 1$, is determined by the antiquark residing in the meson
 of the baryon-pion fluctuation. The resulting effect of the pion cloud 
is thus to add a contribution for negative $\overline x$ and to increase the
 magnitude of the GPDs for positive $\overline x$ with respect to the case of
 the bare proton. In particular, for positive and small $\overline x$ the pion
 cloud contribution as a whole is comparable to that of the bare proton, 
confirming the important role of the sea at small $\overline x.$ 
We also note 
the faster fall off of 
$E^q$ with respect to $H^q$ for $\overline x\to 1$,
showing the decreasing role of the Melosh
 transform to generate angular momentum in $E^q$ with increasing quark 
momentum.

In all cases at $\overline x=0$ the GPDs have a zero. This is due to fact
 that in the overlap integrals the various terms of the proton wavefunction
 are taken at one of their end points. 

The forward limit of the first moment sum rule for the 
spin-averaged GPDs,
is correctly fulfilled. 
For the helicity-flip GPDs, the first moment sum rule 
gives the quark 
anomalous magnetic moments, for which we find the values
$\kappa^u=1.14$ 
and $\kappa^d=-1.03$.
Although these numbers are 5\% and 10\% off the experimental values for 
the up and down quarks, respectively, we
found that
the contribution of the pion cloud 
gives a substantial improvement with respect to the 
results obtained in the model with only valence quarks.

\begin{figure}[t]
\begin{center}
\epsfig{file=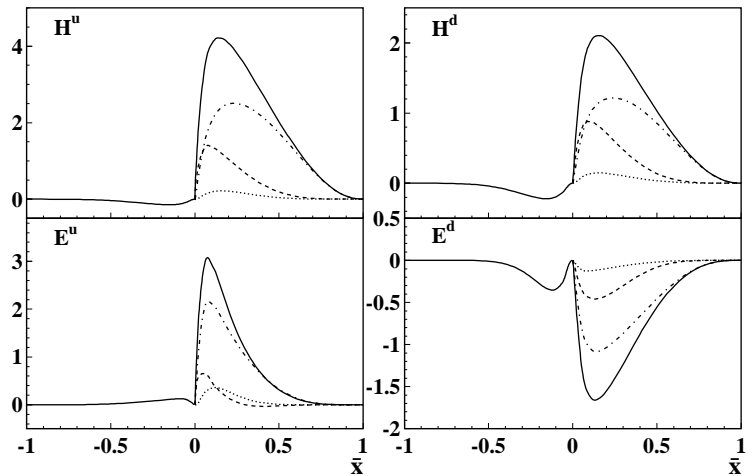,  width=10cm}
\end{center}
\caption{\small The different contributions to the spin-averaged ($H^q$,
 upper panels) and helicity-flip ($E^q$, lower panels) generalized parton 
distributions calculated in the meson-cloud model for flavours $u$ (left 
panels) and $d$ (right panels), at $\xi=0$ and $t=-0.2$ GeV$^2$. Line style 
as in Fig.~\ref{fig:fig1}.
}
\label{fig:fig2}
\end{figure}
Going beyond the forward limit, 
we show 
in Fig.~\ref{fig:fig2} the results for the GPDs
at $\xi=0$ at $t=-0.2$ GeV$^2$. The relative contribution
 of the different components is not modified by switching on the momentum
 transfer $t$, only the overall magnitude is decreased. This is in agreement
 with the common believe that the main part of the $t$ dependence of the GPDs
 is exhibited by their first moments, i.e. by the quark Dirac and Pauli form 
factors.

\begin{figure}[ht]
\begin{center}
\epsfig{file=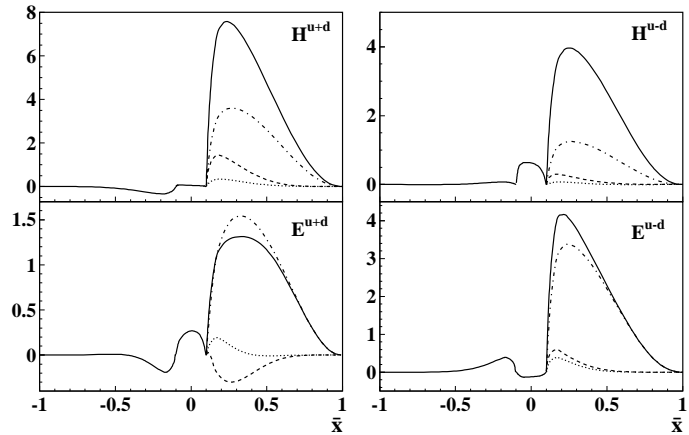,   width=9cm}
\end{center}
\caption{\small Isoscalar ($u+d$, left panels)  and isovector ($u-d$, left
 panels) combinations of the spin-averaged (upper panels) and helicity-flip
 (lower panels) generalized parton distributions calculated in the 
meson-cloud model, at $\xi=0.1$ and $t=-0.2$ GeV$^2$. Line style as in
 Fig.~\ref{fig:fig1}.
}
\label{fig:fig3}
\end{figure}

In  Fig.~\ref{fig:fig3} are plotted 
the isoscalar and isovector combinations of GPDs plotted 
at $\xi=0.1$ and $t=-0.2$ GeV$^2.$ 
We see that GPDs 
in the ERBL region are rather regular functions over the whole range, 
with zeros at the endpoints $\overline  x=\pm\xi$. This result is quite 
different from the oscillatory behaviour predicted by the chiral 
quark-soliton model~\cite{goeke} where the valence contribution of the
 discrete level is a smooth function extending into the ERBL region and
 adding to the sea contribution. Here this is forbidden because the support 
of the valence contribution is limited to the DGLAP region. In addition, 
the transition amplitude between the bare-proton and the baryon-meson 
components vanishes at $\overline x=\pm\xi,$
approaching these points both from the ERBL and DGLAP region.
This generates a discontinuity of the first derivative of GPDs 
at $\overline x=\pm\xi$ which, however, is not in contradiction with 
general principles~\cite{markusthesis}.

Furthermore, in the DGLAP region both for $\overline x\ge\xi$ and $\overline 
x\le -\xi$ no striking difference arises in Fig.~\ref{fig:fig3} for the 
spin-averaged GPDs $H^{u\pm d}$ with respect to the results in the forward
 limit shown in Fig.~\ref{fig:fig1}, while for the helicity-flip GPDs the
 (negative) $d$ contribution coming from the baryon in the $\ket{BM}$ 
component  is responsible for a broader shape at $\overline x\ge\xi$.


\section{Conclusions}
\label{conclusions}

We described a convolution model for the unpolarized GPDs where
the nucleon is viewed by a quark-valence core surrounded by a mesonic cloud.
This meson-cloud model gives the possibility to link GPDs calculated in the 
light-front formalism to the nucleon description in terms of constituent
 quarks including a sea contribution already at a low-energy scale.
Since the contribution in the ERBL region is vanishing in the 
forward limit, it  can not be easily inferred from 
parametrizations
in terms of parton distributions. Therefore, the present calculation
gives new insights to model the off-forward 
features of the GPDs, and can be further used
as a suitable input at the hadronic scale to study the behaviour under 
evolution
at higher scales.

\section*{Acknowledgements}

This research is part of the EU Integrated Infrastructure Initiative
 Hadronphysics Project under contract number RII3-CT-2004-506078.



\end{document}